\documentclass[fleqn,usenatbib]{mnras}

\usepackage{newtxtext,newtxmath}
\usepackage[T1]{fontenc}
\usepackage{ae,aecompl}

\usepackage{siunitx}
\usepackage{graphicx}	% Including figure files
\usepackage{amsmath}	% Advanced maths commands
\usepackage{amssymb}	% Extra maths symbols
\newcommand*\diff{\mathop{}\!\mathrm{d}} %differential d

\newcommand*\vir{\mathrm{vir}}
\title[Filamentary Accretion as Origin of Discs]{Rapid Filamentary Accretion as the Origin of Extended Thin Discs}

\author[M. Kretschmer, O. Agertz and R. Teyssier]{
Michael Kretschmer$^{1}$\thanks{E-mail: michael.kretschmer@physik.uzh.ch},
Oscar Agertz$^{2}$ and
Romain Teyssier$^{1}$
\\
$^{1}$Institute for Computational Science, University of Zurich, Winterthurerstrasse 190, CH-8057 Zurich, Switzerland\\
$^{2}$Lund Observatory, Department of Astronomy and Theoretical Physics, Lund University, Box 43, SE-221 00 Lund, Sweden
}

\date{Accepted XXX. Received YYY; in original form ZZZ}

\pubyear{2020}

\begin{document}
\label{firstpage}
\pagerange{\pageref{firstpage}--\pageref{lastpage}}
\maketitle
%
% Abstract of the paper
\begin{abstract}
Galactic outflows driven by stellar feedback are crucial for explaining the inefficiency of star formation in galaxies. Although strong feedback can promote the formation of galactic discs by limiting star formation at early times and removing low angular momentum gas, it is not understood how the same feedback can result in diverse objects such as elliptical galaxies or razor thin spiral galaxies. We investigate this problem using cosmological zoom-in simulations of two galaxies forming within $10^{12}~\mathrm{M_\odot}$ halos with almost identical mass accretion histories and halo spin parameters.
However, the two resulting galaxies end up with very different bulge-to-disc ratios at $z = 0$.
At $z>1.5$, the two galaxies feature a surface density of star formation $\Sigma_{\rm SFR}\simeq 10~\mathrm{M_\odot}~{\rm yr}^{-1}~{\rm kpc}^{-2}$, leading to strong outflows.
After the last starburst episode, both galaxies feature a dramatic gaseous disc growth from 1~kpc to 5~kpc during 1~Gyr, a decisive event we dub ``the Grand Twirl''.
After this event, the evolutionary tracks diverge strongly, with one galaxy ending up as a bulge-dominated galaxy, whereas the other ends up as a disc-dominated galaxy.
The origins of this dichotomy are the angular momentum of the accreted gas, and whether it adds \emph{constructively} to the initial disc angular momentum.
The build-up of this extended disc leads to a rapid lowering of $\Sigma_{\rm SFR}$ by over two orders of magnitude with $\Sigma_{\rm SFR} \lesssim 0.1~\mathrm{M_\odot}~{\rm yr}^{-1}~{\rm kpc}^{-2}$, in remarkable agreement with what is derived from Milky Way stellar populations. As a consequence, supernovae explosions are spread out and cannot launch galactic outflows anymore, allowing for the persistence of a thin, gently star forming, extended disc.
\end{abstract}
%
%%%%%%%%%%%%%%%%%%%%%%%%%%%%%%%%%%%%%%%%%%%%%%%%%%
%
%%%%%%%%%%%%%%%%% BODY OF PAPER %%%%%%%%%%%%%%%%%%
%
\section{Introduction}
Galaxies have been observed with a variety of shapes and morphologies ranging from ellipticals to spirals.
Although it is yet unclear what are the most important formation mechanisms that determine the final morphology of a galaxy,
it is thought that both large-scale processes, e.g. galaxy interactions and gas accretion, and small-scale processes such as star formation and stellar feedback play important roles. 
The effect of these different mechanisms interact in a non-linear way, making it difficult to disentangle what fundamental physical process really governs their evolution.
Recent models of galaxy formation \citep{2006MNRAS.373.1074S, 2013ApJ...770...25A, 2013MNRAS.429.3068T,2014MNRAS.445..581H,2018MNRAS.475.3283C} have demonstrated that supernovae-driven feedback has a particularly strong impact on the formation of galaxies \citep[for a recent review, see][]{2017ARA&A..55...59N}. 
It is believed that supernovae (SN) explosions release enough energy and momentum in their surrounding medium to eject large quantities of gas from the galaxy into its host halo.
Other feedback mechanisms, such as stellar winds and UV radiation from massive stars contribute as well, although the momentum provided is smaller than the terminal momentum acquired after the Sedov-Taylor phase of SN explosions \citep{2013ApJ...770...25A}.

Furthermore, star formation at small-scales has been shown to be a crucial and delicate ingredient: The right amount of stars formed in the right conditions, at the right location and over the right time-scale will affect the fate of the galaxy. Moreover, since stellar feedback and star formation are closely related to one another, self-regulation mechanisms are likely to emerge and set the characteristic time-scales of the global star formation and gas depletion rates
\citep{2015ApJ...804...18A,2016ApJ...824...79A,2014MNRAS.445..581H, 2015MNRAS.454.2691M,2018MNRAS.480..800H,2017MNRAS.470.4698A,2017ApJ...845..133S}.

The study of galaxies similar to our own Milky Way has proven particularly challenging \citep{2000ApJ...538..477N,2003ApJ...591..499A,2004ApJ...607..688G}.
They are massive galaxies, with deep potential wells, but their overall star formation efficiency, although probably the largest in the entire galaxy population, remains very low, of the order of 20\% of the available baryons in the halo
\citep{1992MNRAS.258P..14P,1998ApJ...503..518F,2000MNRAS.311..793B,2001MNRAS.326..255C,2001MNRAS.326.1228B,2006RPPh...69.3101B,2010MNRAS.404.1111G,2010ApJ...710..903M,2013ApJ...770...57B}.

Recent simulations, including our own \citep{1992ApJ...399L.109K,1993MNRAS.265..271N,1993MNRAS.265..689K,2001ApJ...555L..17T,2006MNRAS.373.1074S,
2008MNRAS.389.1137S,2010ApJ...713..535C,2011MNRAS.410.1391A,2011ApJ...736..134A,2011ApJ...742...76G, 2012MNRAS.423.1726S,2011MNRAS.417..950H,2012MNRAS.424.1275B,2013MNRAS.428..129S,2013ApJ...770...25A,
2014MNRAS.442.1545C,2015ApJ...804...18A,2015MNRAS.450..504M,2014MNRAS.445..581H,2013MNRAS.436.3031V}, have shown that obtaining such a low gas-to-star conversion efficiency is made possible by vigorous feedback-driven galactic outflows. In terms of Milky Way-like galaxies, outflows are thought to be especially important in their progenitors at high redshift \citep[for a recent review, see][]{2015ARA&A..53...51S}, an epoch when they were highly turbulent, gas rich thick discs featuring massive star forming clumps \citep{2008ApJ...688...67E, 2013ApJ...768...74T, 2011ApJ...733..101G}.
The violent character of this early epoch of galaxy formation is a key feature in recent numerical models, and is supported by observations of distant galaxies \citep{1996ApJ...462L..17S,1999ApJ...519....1S,2001ApJ...555..301M,2004ApJ...604L..21E,2006Natur.442..786G}.

On the contrary, present day Milky Way-mass galaxies are, broadly, either 1) quiescently star forming disc galaxies, featuring thin gaseous discs with low levels of turbulence and no sign of strong galactic outflows, or 2) gas-poor, almost non-star-forming (quenched) elliptical galaxies \citep{2009ApJS..182..216K}.
How galaxies transition between the early, turbulent, outflow-driven, evolution at $z>1$, to settling into large and thin galactic discs, seemingly undisturbed by vigorous feedback, is not yet understood.
In the Milky Way, there is observational evidence that this transition may have been rapid \citep[several hundreds of Myr, see][]{2014ApJ...789L..30L}, posing a challenge for galaxy formation models.

Strong outflows with gas velocities larger than the escape velocity of their host halo have been observed for example by \cite{2003RMxAC..17...47H} \citep[see also the recent review on observations of galactic winds by][]{2018Galax...6..138R}, with
outflow velocities correlating with the surface density of star formation rate $\Sigma_{\rm SFR}$ \citep[][and references therein]{2016ApJ...822....9H, 2019ApJ...873..122D}.
Interestingly, strong galactic outflows occur only above a certain threshold $\Sigma_{\rm SFR}$. In local galaxies this was found to be $\Sigma_{\rm SFR} > 0.1~\mathrm{M_\odot yr^{-1}~kpc^{-2}}$ whereas at higher redshift this value was found to be larger, namely $\Sigma_{\rm SFR} > 0.5 - 1~\mathrm{M_\odot yr^{-1}~kpc^{-2}}$ \citep{2012ApJ...761...43N,2019ApJ...875...21F}.

Such a phenomenon is found in theoretical studies using stratified box simulations \citep{2015MNRAS.454..238W, 2016MNRAS.456.3432G} where strong outflows only occur when the number surface density of SN explosions in the disc exceeds a similar threshold -- clustered SN events appear to be a key ingredient for powering strong outflows.

While strong feedback is a promising way to explain the low efficiency of galaxy and star formation, the same strong feedback can also destroy the thin galactic discs, leading to nonphysical thick and turbulent discs at low redshifts \citep{2007MNRAS.374.1479G,2011MNRAS.410.1391A,2013A&A...554A..47G,2013MNRAS.436..625S,2014MNRAS.444.2837R,2020MNRAS.492.1385K}. Invoking strong feedback alone is therefore not sufficient. We need an additional mechanism to suppress feedback once large and thin discs have formed at low redshifts. 

This paper aims at studying in detail the transition between high and low redshift galaxy formation, i.e. between the early feedback dominated epoch to the present-day quiescent regime.
We believe that this transition is the key to our understanding of the settling and existence of extended discs and the morphological differentiation between disc-dominated and bulge-dominated galaxies, as well as the observed epoch of ``disc settling'' ($z\lesssim 1$), where the neutral gas velocity dispersion in star forming galaxies is found to decrease over time \citep{2012ApJ...758..106K}.

In this work we investigate this problem using cosmological zoom-in simulations of two galaxies forming in $10^{12}~M_\odot$ halos with almost identical mass growth histories. After a starburst episode triggered by the last major merger at $z\sim 1.5$, the two galaxies similar evolutionary paths diverge, with one galaxy ending up as a compact bulge-dominated galaxy, whereas the other evolves to a disc-dominated galaxy
analogue to our Milky Way.
The reason for these different outcomes is the angular momentum of the accreted gas, and whether it adds \emph{destructively} or \emph{constructively} to the forming disc.

We have structured this paper as follows: In \autoref{section:methods} we present our numerical methodology and summarize the properties of the subgrid models we have used. In \autoref{section:results} we present our results, focusing on the description of the last major starbursts around $z \simeq 1.5$, followed by the fast assembly of a large gaseous discs leading to the formation of a disc-dominated system analogue to our Milky Way. We describe how the formation of such a large disc naturally explains why supernovae feedback stops to be efficient, leading to the maintenance of a quiescent regime.
We systematically compare the properties of our disc-dominated system to the other simulation leading to a bugle-dominated galaxy. 
We finally discuss the implications of our work in \autoref{section:discussion} 
and conclude in \autoref{section:conclusions}.

\section{Methods}
\label{section:methods}

\subsection{Halo selection and initial conditions}

Our study is based on two cosmological zoom-in simulations performed with the adaptive mesh refinement (AMR) code \textsc{RAMSES} \citep{2002A&A...385..337T}.
Our methodology is as follows: We first ran an $N$-body simulation with $512^3$ dark matter particles in a periodic box of size 25~$h^{-1}$Mpc. 
From this volume we selected, at $z=0$, two halos with virial masses $M_\mathrm{vir}=1.1 \times 10^{12}\mathrm{M_\odot}$ and $M_\mathrm{vir}=1.4 \times 10^{12} \mathrm{M_\odot}$ 
in relative isolation; no halo more massive than half $M_\mathrm{vir}$ within 5 virial radii exists. Note that the virial mass is calculated here using a spherical over-density according to the definition of \cite{1998ApJ...495...80B}.

We then generated higher resolution initial conditions around the selected halos using the MUSIC code \citep{2011MNRAS.415.2101H}, with an initial hierarchy of concentric grids from $\ell_{\rm min}=7$, corresponding to a coarse grid resolution of $128^3$ covering the entire periodic box, to $\ell_{\rm max,ini}=11$, corresponding to an effective initial resolution of $2048^3$. This yields  a dark matter particle mass of $m_{\rm dm}=2.0 \times 10^{5}$M$_\odot$ and an initial baryonic  mass resolution of $m_{\rm bar}=2.9 \times 10^{4}$M$_\odot$.

We subsequently carried out our final simulations including gas and galaxy formation physics, as summarized below. The maximum resolution was set to $\ell_{\rm max}=19$ at $z=0$, with refinement levels progressively released to enforce a quasi-constant physical resolution, where the smallest cells have sizes $\Delta x_{\rm min}=55$pc. The adopted refinement criterion is the traditional quasi-Lagrangian approach, namely cells are individually refined when more than 8 dark matter particles are present or when the total baryonic mass (gas and stars) exceeds $8\times m_{\rm bar}$.
Only the Lagrangian volume corresponding to twice the final virial radius of the halo was adaptively refined, the rest of the box being kept at a fixed, coarser resolution to provide the proper tidal field. 

We intentionally picked halos without any major mergers at $z<1$ and with overall similar mass accretion histories. 
The halos feature a strong merger-induced starburst at $z=1.4$ and $z=2.0$ respectively, followed by quiescent accretions history until $z=0$. The final dark matter halo spin parameters, as defined in \cite{2001ApJ...555..240B}, are very similar for both halos, namely $\lambda = 0.015$ and $\lambda = 0.014$. 
This fact is central to our study, as the resulting galaxies end up with very different bulge-to-disc ratios at $z=0$, indicating that 
the dark matter halo spin is not a robust predictor for the final galaxy angular momentum and size \citep[e.g.][]{2019MNRAS.488.4801J}.

\subsection{Galaxy Formation Physics}

For our star formation recipe, we adopt a traditional Schmidt law, for which the star formation rate density is given by 
\begin{equation}
\dot{\rho}_\star=\epsilon_{\rm ff} \frac{\rho}{t_\mathrm{ff}}
\end{equation}
where $\rho$ is the density of the gas, $t_\mathrm{ff}=\sqrt{{3\pi}/{(32 G \rho)}}$ is its free-fall time and $\epsilon_{\rm ff}$ is the star formation efficiency per free-fall time.
Traditionally, $\epsilon_{\rm ff}$ is chosen to be a constant, close to $1\%$, and star formation
is allowed only in gas cells above a prescribed density threshold and below a prescribed temperature threshold.
This is motivated by observations of inefficient star formation on galactic kpc scales \citep{2008AJ....136.2846B} as well as in Milky Way giant molecular clouds (GMCs) \citep{2007ApJ...654..304K}, albeit with a 1 dex spread \citep{2011ApJ...729..133M,2016ApJ...833..229L,2019MNRAS.486.5482G}.

In this paper, we use a novel approach for which $\epsilon_{\rm ff}$ depends on the turbulent state of the gas following the so-called \textit{multi-freefall} approach
\citep{2012ApJ...761..156F,2016ApJ...826..200S,2020MNRAS.492.1385K}.
Indeed, in a turbulent medium, like the interstellar medium (ISM), the gas density distribution is well described by a log-normal probability distribution function (PDF):
\begin{equation}
p(s)=\frac{1}{\sqrt{2\pi \sigma_s^2}}\exp {\frac{(s-\overline{s})^2}{2\sigma_s^2}},
\end{equation}
 where $s=\ln \rho / \rho_0$, $\sigma_s$ is the variance of $s$, $\rho$ is the gas density and $\rho_0$ the mean density.
 
Gas with densities larger than a critical density $s_\mathrm{crit}$, corresponding to a collapse criterion, is converted into stars.
The star-formation rate per free fall time $\epsilon_\mathrm{ff}$ is obtained by integrating the PDF weighted by the density and the inverse free-fall time above $s_\mathrm{crit}$ \citep{2011ApJ...743L..29H,2012ApJ...761..156F}:
\begin{equation}
\begin{split}
\epsilon_\mathrm{{ff}} &= \frac{\epsilon}{\phi_t}\int_{s_\mathrm{crit}}^{\infty}\frac{t_\mathrm{ff}(\rho_0)}{t_\mathrm{ff}(\rho)}\frac{\rho}{\rho_0} p(s) \diff s \\
&= \frac{\epsilon}{2\phi_t} \exp \left( \frac{3}{8} \sigma_s^2\right)\left[ 1 + \mathrm{erf}\left(\frac{\sigma_s^2-s_\mathrm{crit}}{\sqrt{2 \sigma_s^2}}\right) \right].
\label{eq:sfr-ff}
\end{split}
\end{equation}
We compute $s_\mathrm{crit}$ using the model of \cite{2005ApJ...630..250K} for which $s_\mathrm{crit}$ is estimated by requiring the
virial parameter of the gas to be less than 1. We slightly modify the original model to account for subsonic and transonic cases such that for $\mathcal{M}<1$ we require that the entire computational cell becomes gravitational unstable.
With this, we can use the modified model for both regime $\mathcal{M} \leq 1$ and $\mathcal{M} \geq 1$.
The obtained critical density for star formation is
\begin{equation}
s_\mathrm{crit} = \ln \left[ \alpha_\mathrm{vir} \left( 1 +  \frac{2 \mathcal{M}^4}{1+\mathcal{M}^2} \right)\right],
\end{equation}
where
\begin{equation}
\alpha_{\rm vir} = \frac{5\sigma^2}{3 G \rho \Delta x^2}
\end{equation}
is the local virial number which can be interpreted as an estimator for the local stability, $\Delta x$ is the cell size and $\sigma$ is the turbulent 1D velocity dispersion which is computed by the subgrid scale (SGS) turbulent energy model \citep[see][for more details]{2006A&A...450..265S,2016ApJ...826..200S,2020MNRAS.492.1385K}.
As a consequence, the efficiency is not a constant anymore but a function of the state of the gas in a computational cell which is characterized through the local virial parameter $\alpha_\mathrm{vir}$ and the turbulent Mach number $\mathcal{M}$. This model therefore has two star formation channels, either $\alpha_\mathrm{vir}<1$ where the whole cell is collapsing under gravity or if $\mathcal{M}$ is large, such that large density fluctuations caused by supersonic turbulence occur. 

\begin{figure*}
    \centering
    \includegraphics[width=\linewidth]{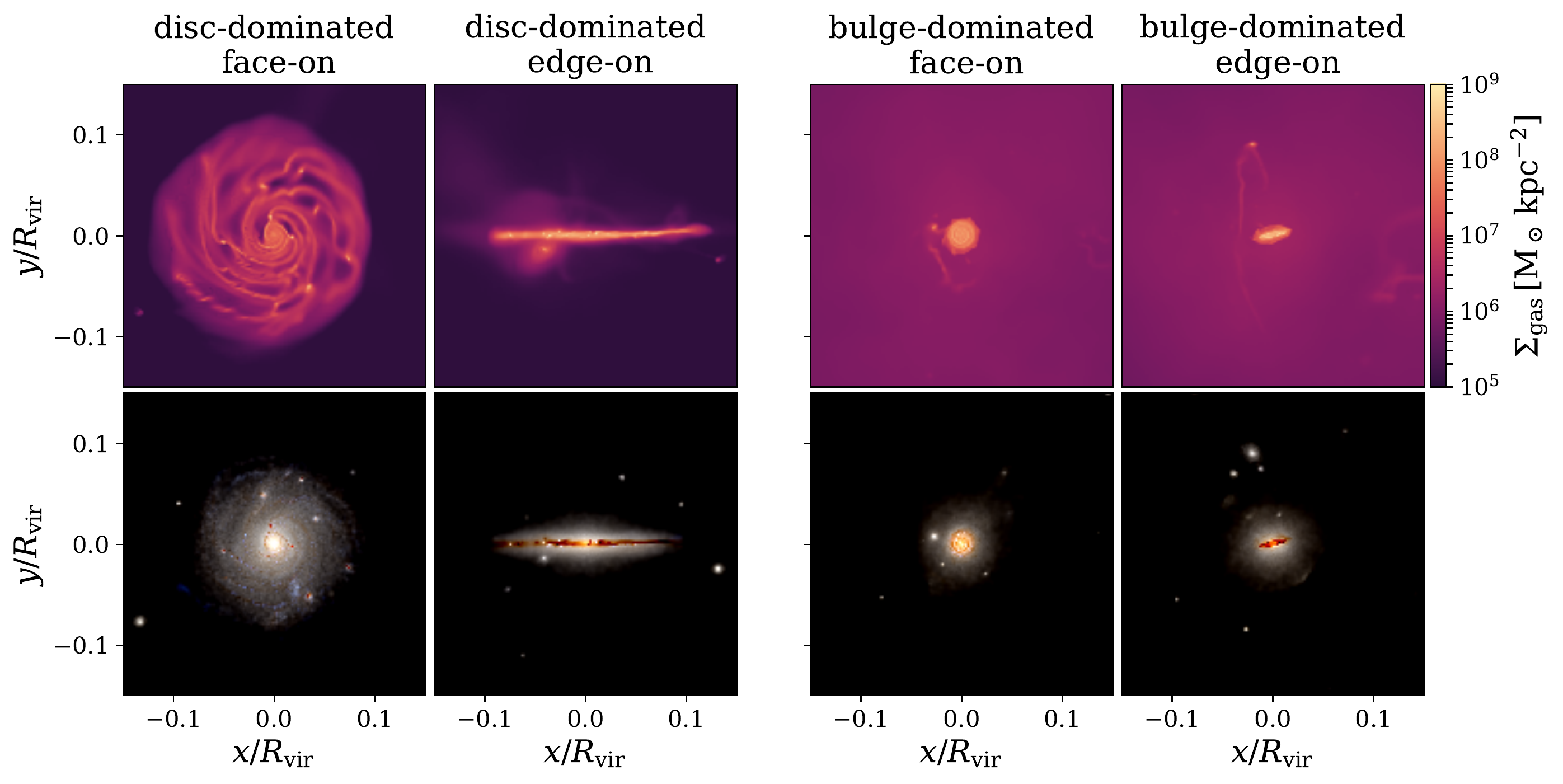}
    \caption{Projected maps of the two galaxies at $z=0$.
    Left two columns are the disc-dominated galaxy face-on and edge-on. Right two columns are the bulge-dominated galaxy face-on and edge-on. The top row shows the gas surface density and the bottom row shows true colour images with dust absorption allowing us to render both stars and gas.}
    \label{fig:vis}
\end{figure*}

Stellar feedback is implemented as both thermal and momentum injection from supernovae as described in \cite{2013ApJ...770...25A, 2020MNRAS.492.1385K}. A supernova remnant evolves from the energy conserving (adiabatic) phase to the momentum conserving phase. The transition happens when radiation losses become important, or when the shock radius reaches the so-called cooling radius $R_{\rm cool}$. Resolving this scale by at least three computational cells is crucial in order for the adiabatic phase to inject the correct amount of momentum into the surrounding gas \citep[][]{2015ApJ...802...99K}. If the cooling radius is unresolved, which can occurs at high gas densities, the thermal energy of the supernova is spuriously radiated away. In this case, in addition to the thermal energy injection, we also inject the predicted amount of terminal momentum ($P_\mathrm{SN}$), as computed by  \cite{2015MNRAS.450..504M}. 

In this work, we use for the cooling radius and the terminal momentum the values computed by \cite{2015MNRAS.450..504M} for an \textit{in-homogeneous medium} (as opposed to the values for an \textit{homogeneous medium}).
Specifically, we adopt
\begin{equation}
R_\mathrm{cool} = 6.3\, \mathrm{pc}\left(\frac{Z}{Z_\odot}\right)^{-0.05}  \left( \frac{n_\mathrm{H}}{100\,\mathrm{cm}^{-3}}\right)^{-0.42},
\end{equation}
where $n_\mathrm{H}$ is the gas density of the cell where the star is exploding, and 
\begin{equation}
P_\mathrm{SN} = 1.1 \times 10^5  \mathrm{km\,s^{-1}M_\odot} \left(\frac{Z}{Z_\odot}\right)^{-0.114}\left(\frac{n_\mathrm{H}}{100\,\mathrm{cm}^{-3}}\right)^{-0.19}.
\end{equation}
The injected thermal energy is set to $E_\mathrm{SN}=10^{51} \mathrm{erg}$.  We resolve individual supernovae explosions in time, with each star particle triggering multiple supernovae explosion, spread randomly between $3$ and $20$ Myrs after the birth of the star particle according to the age distribution of massive stars \citep{2020MNRAS.492.1385K}. Each supernova injects a mass of metals set to 10\% of the progenitor mass. The metallicity is advected by the fluid as a passive scalar.

Gas cooling and heating are implemented where equilibrium chemistry for Hydrogen and Helium is assumed \citep{1996ApJS..105...19K} as well as metal cooling at both low and high temperature. Additionally heating by the UV background and its self-shielding is taken into account where the self-shielding density is assumed to be $n_\mathrm{H}=10^{-2} \mathrm{H/cc}$ \citep{2010ApJ...724..244A}. 
Heating from the cosmic UV background is turned on at $z_\mathrm{reion}=10$ \citep{1996ApJ...461...20H}.

Subgrid physics always remains highly tentative but we emphasise that the robustness of our models have been demonstrated in \cite{2020MNRAS.492.1385K} as well as in other independent studies \citep{2016ApJ...826..200S,2018MNRAS.477.1578H,2020MNRAS.495..199G,2020arXiv200406008N}. Even more important, gas accretion is a more robust feature of our model than the details of our subgrid physics.
The behaviour of the CGM gas and its AM, is less dependent on the adopted subgrid physics. This is because gas at larger scales is more predictable than star formation and its feedback inside galaxies (apart from the CGM-wind interaction).

\section{Results}
\label{section:results}
We now present our results for our two selected halos. 
Recall that both halos have a similar accretion history, with an early assembly and a quiet late evolution.
The final total masses at $z=0$ ($M_\mathrm{vir}=1.1 \times 10^{12}\mathrm{M_\odot}$ and $M_\mathrm{vir}=1.4 \times 10^{12} \mathrm{M_\odot}$) are both similar to the estimated halo mass of the Milky Way \citep[e.g.][]{2019MNRAS.484.5453C,2020MNRAS.494.4291C,2020SCPMA..63j9801W}.
Despite the fact that the two halos also have identical halo spins, the two emerging galaxies have very different properties. From \autoref{fig:vis}, it is immediately apparent that at $z=0$ one simulation produced a large, extended thin gas disc. In contrast, the other galaxy contains a small nuclear gas disc within a spherically distributed stellar system. We classify the first galaxy as disc-dominated and the second as bulge-dominated. This visual classification will be confirmed later using a more detailed kinematic analysis (\autoref{section:eccentricity}). 

\begin{figure}
    \centering
    \includegraphics[width=\columnwidth]{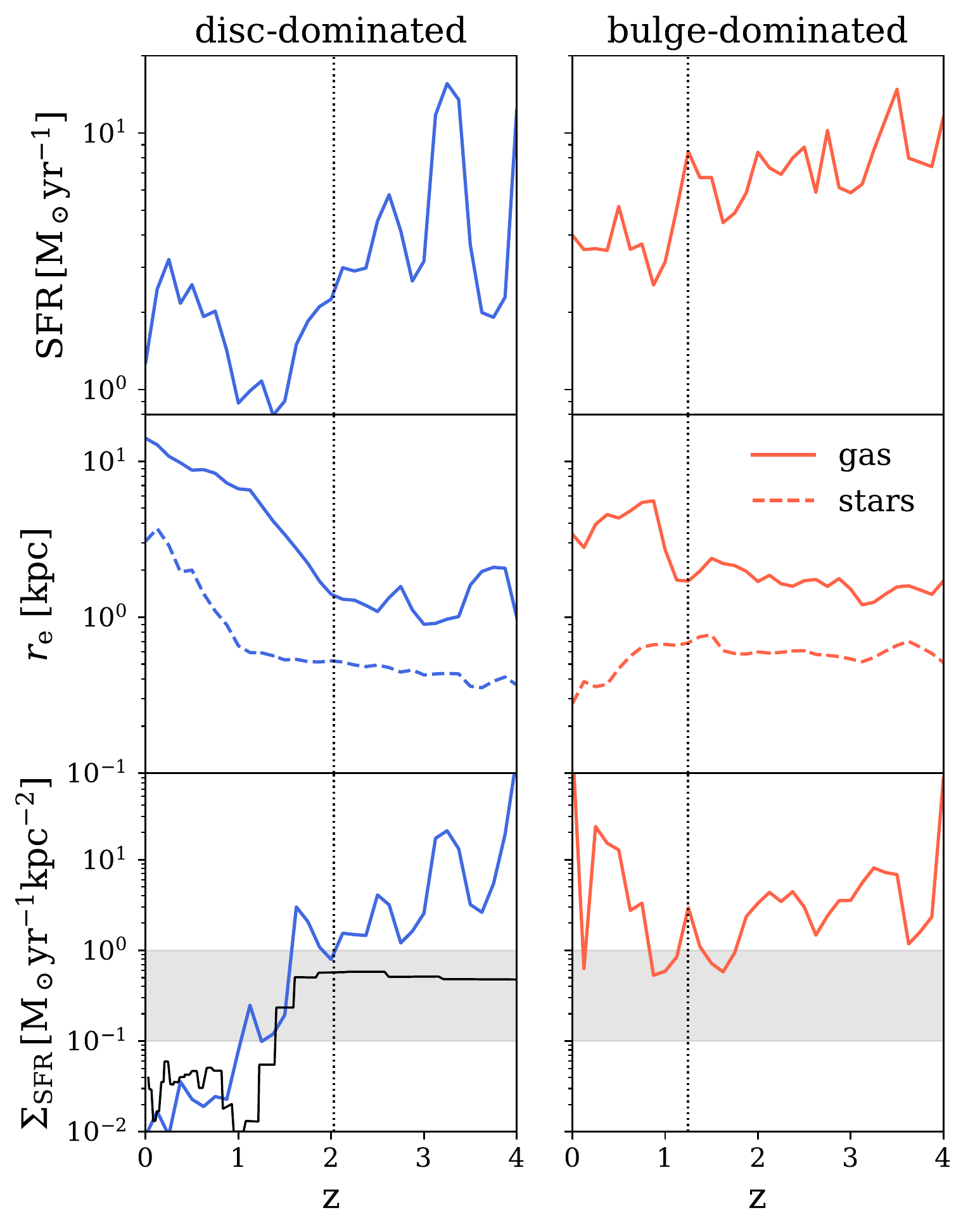}
    \caption{From top to bottom: Evolution of SFR, half-mass radius $r_e$ of the gas and stars, and 
    evolution of star-formation intensity over redshift for both galaxies.
    Left: The galaxy that will end up disc-dominated. Right: The galaxy that will end up bulge-dominated galaxy.
    The dotted vertical lines indicate the moment of the last starbursts in each galaxy.
    The shaded area indicates the threshold for starburst-driven outflows (see text).
    The thin black line in the bottom left plot is the evolution for the Milky Way obtained by \protect\cite{2014ApJ...789L..30L}.
    }
    \label{fig:sfr_acc}
\end{figure}
\begin{figure}
    \centering
    \includegraphics[width=\columnwidth]{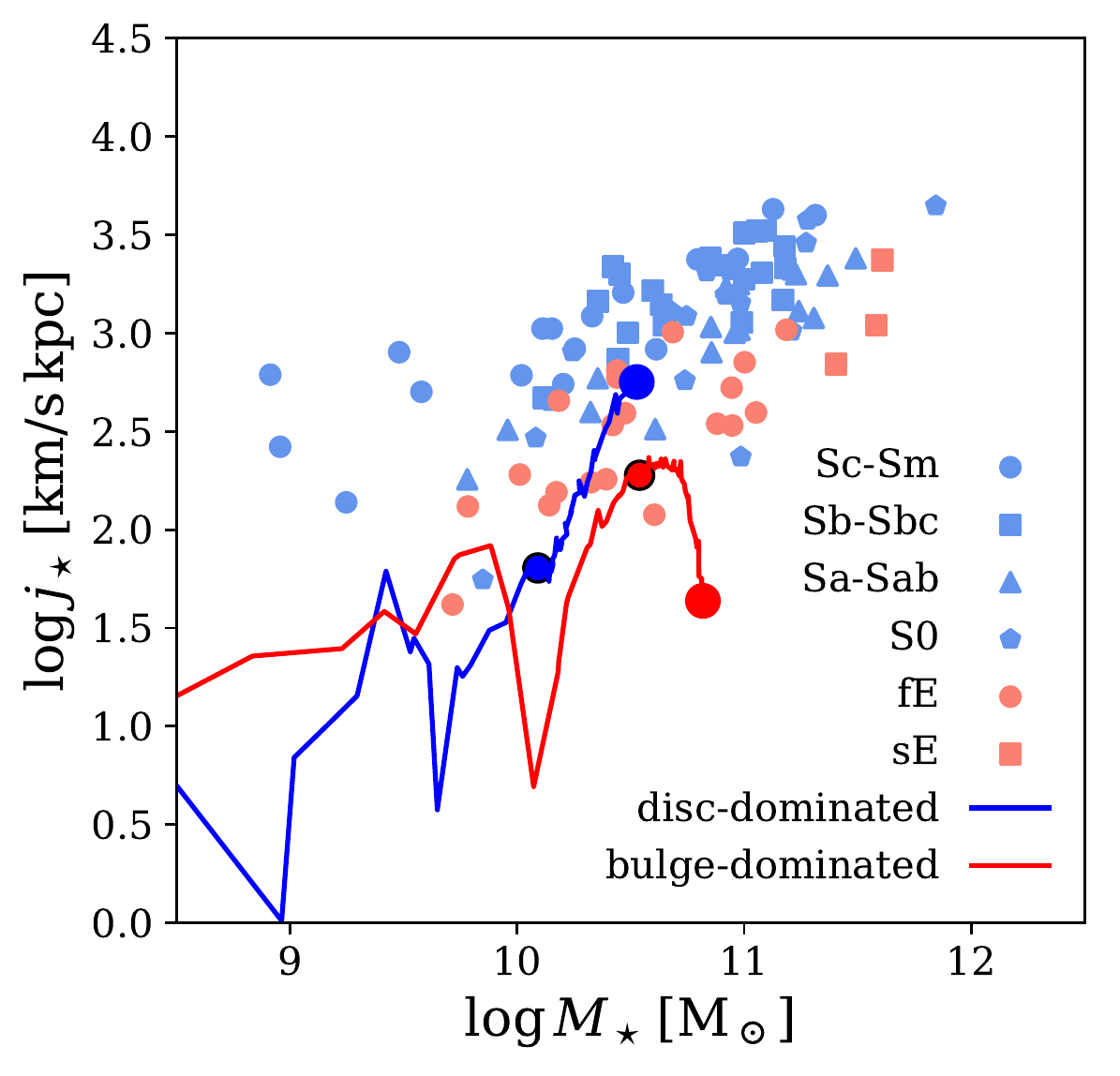}
    \caption{Evolution of the specific angular momentum of the stars for the two galaxies as a function of their stellar mass. Blue is the galaxy that will end up as a disc-dominated galaxy at $z=0$ and red is the bulge-dominated. The small circle in the same color of the line is the moment of the last starburst. The big circle is the point at $z=0$. Data points are taken from \protect\cite{2013ApJ...769L..26F}, where circles are ellipticals and others are spirals.}
    \label{fig:am_evo}
\end{figure}
\begin{figure*}
    \centering
        \includegraphics[width=0.9\linewidth]{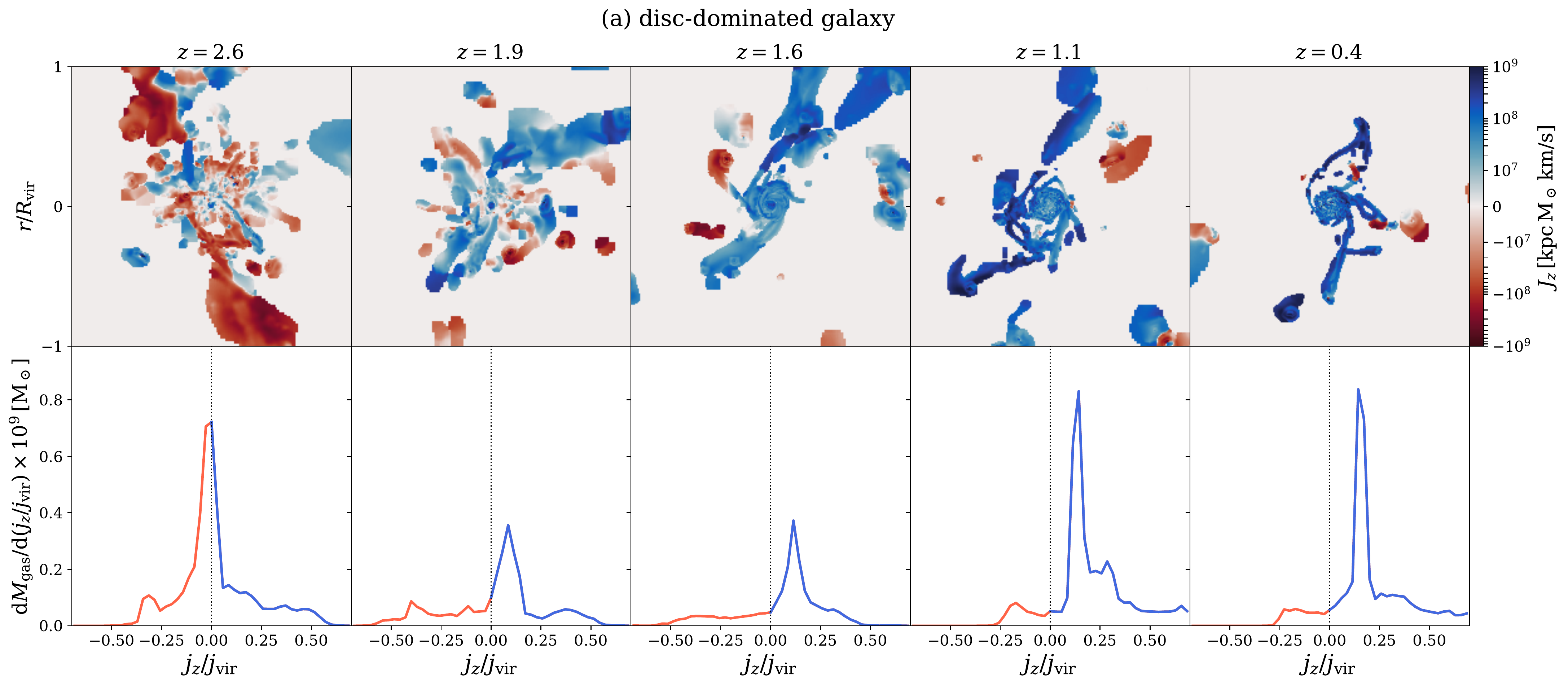}
        \includegraphics[width=0.9\linewidth]{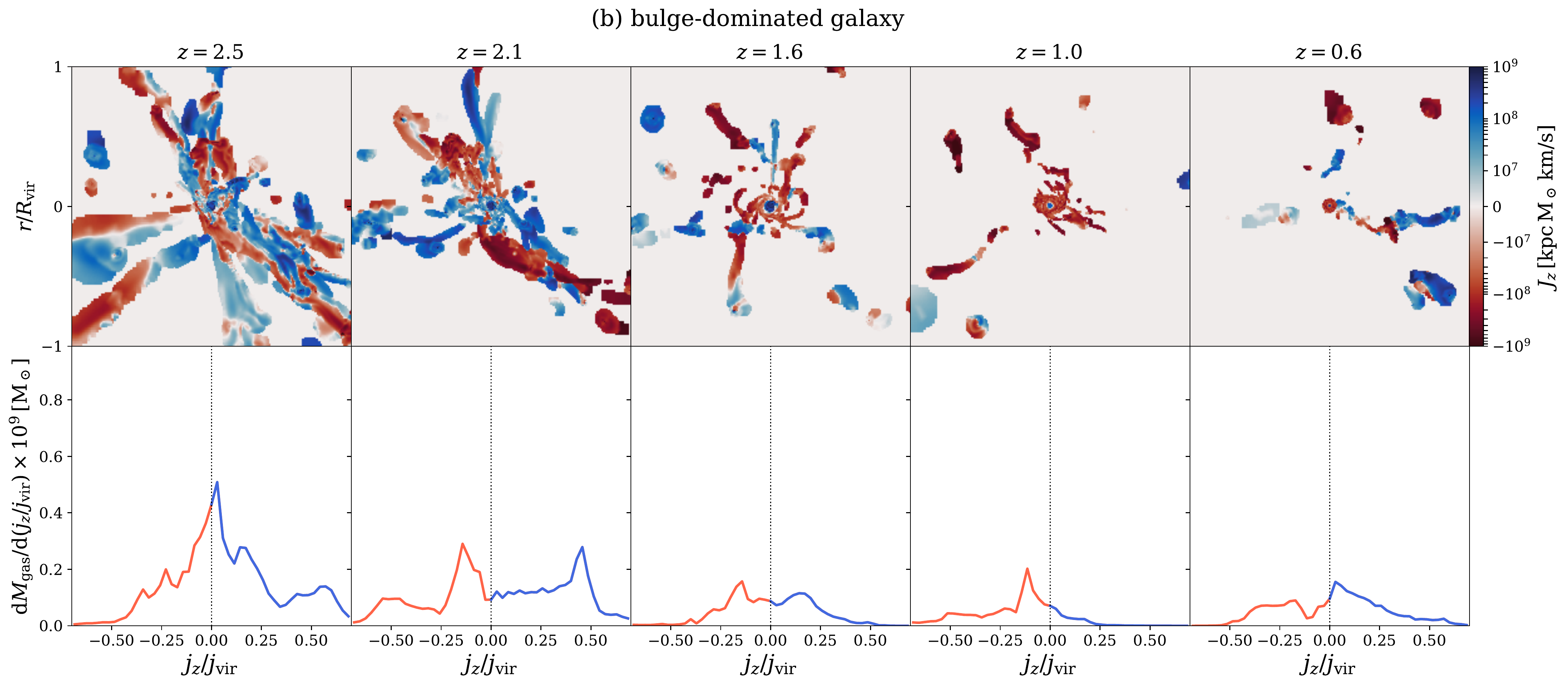}
    \caption{
    Evolution of the mass-weighted cold gas angular momentum (AM) in the virial radius of the disc-dominated galaxy (a) and the bulge-dominated galaxy (b).
    Plots are done at $z\sim 2.5, 2.0, 1.5, 1.0, 0.5$.
    \textbf{Top rows:}
    Mass-weighted map of cold gas angular momentum inside the virial radius.
    Colours indicate the sign of the angular momentum where red corresponds to counter-rotating gas and blue to co-rotating.
    For the disc-dominated galaxy, most of the cold gas is joining the disc constructively. An extended disc of cold gas is formed.
    In contrast, for the bulge-dominated galaxy, most of the cold gas is counter-rotating with respect to the disc. This causes the galactic disc to shrink.
    \textbf{Bottom rows:}
    Mass-weighted distribution of specific angular momentum of infalling cold gas measured between $0.1 < r/R_\mathrm{vir} < 1.0$ normalised by the specific virial AM of a given snapshot. $j_z$ is measured with respect to the disc AM inside $0.1 R_\mathrm{vir}$. 
    For the disc-dominated galaxy, most infalling cold gas has positive $j_z/j_\mathrm{vir}$ resulting in constrictive AM.
    For the bulge-dominated galaxy, the distribution shows positive and negative contributions where sometimes the destructive part is dominating.
    Infalling cold gas therefore adds destructive AM to the disc.
    }
    \label{fig:am}
\end{figure*}
\subsection{Star Formation History and Gas disc Evolution}
It is today established that strong feedback is required, especially at high redshift, to obtain a realistic amount of stars in
present day galaxies \citep[][]{2017ARA&A..55...59N}. In our simulations, the early evolution is dominated by strong outflows driven by
intense starbursts. \autoref{fig:sfr_acc} shows the evolution of the instantaneous Star Formation Rate ($\mathrm{SFR}$) within 
10\% $R_{\rm vir}$ for our two galaxies. They both experience a sequence of starbursts with $\mathrm{SFR\simeq 10~M_\odot yr^{-1}}$ until redshift 1.5. 
The dashed lines indicate the moment of the last starbursts in each galaxy (defined as the last peak in the SFR) which is accompanied by vigorous outflows with mass loss rates $\dot{M}_\mathrm{out}\sim 15-25~\mathrm{M_\odot yr}^{-1}$. It is after this event that the epoch of disc formation truly begins.

\autoref{fig:sfr_acc} also shows the time evolution of the effective radius $r_{\rm e}$ (half-mass radius) of the cold $(T<5\times 10^4\mathrm{K})$ gas disc inside $0.1 R_\vir$. After the last starburst, both galaxies rapidly develop extended discs
\citep[see also][]{2020arXiv200606008A,2020arXiv200606011R,2020arXiv200606012R, 2020MNRAS.496.5372D}.
More precisely, both galaxies feature $r_{\rm e} \simeq 1 $~kpc before the last starburst, and end up with $r_{\rm e}\sim 5$~kpc after only a few Gyr. In fact, the half-mass radius doubles in only $750$~Myr. We call this event of spectacularly fast disc growth the {\it Grand Twirl}. In the case of the disc-dominated galaxy, $r_{\rm e}$ continues to grow, reaching $10$~kpc at $z=0$, while the bulge-dominated galaxy shrinks back to $r_{\rm e}\sim 2$~kpc.

We also show the evolution of the effective radius of the stars $r_{e\star}$ in $0.1 R_\vir$. For the disc-dominated system, $r_{e\star}$ growth after the Grant Twirl significantly from $\sim 0.5$kpc to $\sim 3$kpc at $z=0$ similar to the value of the Milky Way of $4-5$kpc \citep{2011MNRAS.414.2446M,2020MNRAS.494.4291C}. 

Interestingly, during this Grand Twirl, the instantaneous SFR drops significantly in both galaxies.
This is due to quenching by gas depletion and outflows following the merger driven starburst \citep[e.g.][]{2014MNRAS.442L..33R} which decreases the gas surface densities across the entire (rapidly expanding) discs. After this short period, gas re-accretion leads to the SFR increasing to 3-5~$\mathrm{M_\odot yr}^{-1}$ in both galaxies, albeit with no more strong outflows in the disc-dominated
case. Indeed, for this Milky Way-like galaxy, the Grand Twirl marks the transition between the high-redshift, outflow dominated regime and the low-redshift, quiescent, disc-dominated regime.
At $z=0$ the SFR of the disc-dominated galaxy is $\mathrm{SFR} = 1.3 \mathrm{M_\odot yr}^{-1}$ which is in good agreement with values obtained for the MW \citep[see e.g.][where $\mathrm{SFR} \sim 1.7 \mathrm{M_\odot yr}^{-1}$]{2015ApJ...806...96L}.
We note that the star formation history obtained from the simulated galaxies is to some extend determined by the chosen sub-grid models. However, possible deviations from observed properties are not detrimental to our results since the evolution of the angular momentum of the gas is mostly dependent on physical processes outside the disk at larger scales.
In the following sections we will outline why the galaxies diverge in terms of their overall properties and why outflows shut down in the disc-dominated galaxy.

\subsection{Stellar Disc Evolution}

In order to outline the differences between the two galaxies, we now analyse the evolution of the stellar discs.
For this, we compute the specific angular momentum of all stars within $0.1 R_\vir$ and compare its time evolution 
to observations from \cite{2013ApJ...769L..26F} in \autoref{fig:am_evo}.
The evolution of the angular momenta for the two galaxies is very similar, with $j_\star \simeq 100-200~{\rm km/s~kpc}$ up to the time of the last starburst, which is marked as a small outlined circle in the figure. During this early epoch, $j_\star$ increases roughly in proportion to the stellar mass of the system. 

After the last starburst and the Grand Twirl, on the other hand, the two evolutionary tracks diverge:
For the disc-dominated galaxy $j_\star$ increases by more than one order of magnitude and reaches the regime of observed spiral galaxies \citep[for similar examples, see][]{2016ApJ...824...79A,2017ApJ...835..289S}.
The specific angular momentum here grows as $j_\star\propto M_\star^2$ \citep[see also][]{2014A&A...567A..47P}, which is steeper than the expected relation for dark matter halos \citep[$j \propto M_{\rm vir}^{2/3}$,][]{1980MNRAS.193..189F}.
We see that the specific angular momentum increased by a factor of $\sim 9$ and the total stellar AM increases after the Grand Twirl at $z\sim2$ by a remarkable factor of $\sim 24$ from $7.9 \times 10^{11} \mathrm{M_\odot km s^{-1} kpc}$ to $1.9 \times 10^{13} \mathrm{M_\odot km s^{-1} kpc}$. The growth by such a remarkably large factor is in good agreement with recent studies \citep{2017MNRAS.467.3140S,2019A&A...621L...6M,2020MNRAS.491L..51P,2020MNRAS.495L..42R}.
In contrast, the specific angular momentum of the bulge-dominated galaxy decreases by one order of magnitude and ends up in the observational regime of elliptical galaxies. 
This classification in the $j_\star$ vs $M_\star$ plane is in clear agreement with the visual classification in \autoref{fig:vis}.

\subsection{Cold Gas Accretion and Angular Momentum Build-Up}

To understand the origin of the diverging evolutionary tracks in the $j_\star$/$M_\star$ plane, 
we next turn to the angular momentum of the cold gas $(T<5\times 10^4\mathrm{K})$ within the virial radius, but outside $0.1 R_\vir$. \autoref{fig:am} shows face-on maps of the mass-weighted distribution of $J_z$ at different epochs. 
The reference frame of the galaxy is computed using the total angular momentum of the cold gas inside $0.1 R_\vir$.

At high redshift $(z\simeq2.5)$ both galaxies show significant amounts of cold gas both counter (red)- and co-rotating (blue). 
However, going forward in time, it is apparent that for the disc-dominated galaxy, most of the cold gas is co-rotating. 
In particular, we see co-rotating gas streams, originating from the cosmic web, directly merging with the galactic disc. 
We also see the size of the galactic disc steadily increasing with time, with all of the disc's gas settled into co-rotation.

In contrast, the halo of the bulge-dominated galaxy is filled with mostly counter-rotating gas originating at large radii and merging with the disc.  Quite strikingly, the Grand Twirl in this scenario, while producing an outer extended disc, is entirely \emph{counter-rotating} in relation to the inner disc (see panels for $z\le 1.6$). This negative angular momentum gas is slowly diminishing the net angular momentum of the central galactic disc, until there is only a tiny cold gas disc left with a counter-rotating outer ring.

To quantify the above points, the distribution of the cold halo gas angular momentum is shown below each image in the bottom rows of \autoref{fig:am} 
in units of $j_{\rm vir} = V_{\rm vir} R_{\rm vir}$. 
At early times $(z\simeq2.5)$, the distribution is dominated by almost radially in-falling gas.
For the disc-dominated galaxy, after the Grand Twirl, the AM distribution shows a clear peak 
around $j_z / j_{\rm vir} \simeq + 0.1$. For reference, recall that the spin parameter of the halo is more than five times smaller than this
\citep[see also][where a similar factor was found in a study of independent hydrodynamical simulations]{2017ApJ...843...47S}.
This peak persists and signals a consistent feeding of the galactic disc with fresh gas and a {\it constructive} buildup of angular momentum.
For the bulge-dominated galaxy, however, we find a peak in the AM distribution around $j_z / j_{\rm vir} \simeq - 0.2$. This stream of gas contributes {\it destructively} to the AM of the central galaxy disc, explaining why the disc size shrinks at late time.

In summary, the disc morphology at $z=0$ comes from cold gas being accreted constructively relative to the initial disc AM, 
while the bulge-dominated morphology (with a small surviving nuclear disc) originates from late accretion with a destructive impact on the disc AM. This gas is a mixture of pristine gas accreted from the cosmic web and metal-enriched gas recycled from previous galactic outflows.

\subsection{Kinematics Analysis of the Stellar discs}
\label{section:eccentricity}

We next analyse the angular momentum distributions of the stars in the two galaxies.
For this we use the circularity parameter $\epsilon$, defined as $\epsilon=j_z/j_c(E)$, 
where $j_z$ is the $z$-component of the specific angular momentum and $j_c(E)$ is the specific angular momentum for a pure circular orbit at the energy $E$ of the corresponding star particle. For a prograde circular orbit in the midplane of the disc, $\epsilon=+1$, and $\epsilon=-1$ for a retrograde circular orbit. For inclined or eccentric orbits, $\epsilon$ will be $-1 < \epsilon < 1$.
For a classical bulge, for example, we expect a symmetric distribution around $\epsilon=0$.

\begin{figure}
    \centering
    \includegraphics[width=\columnwidth]{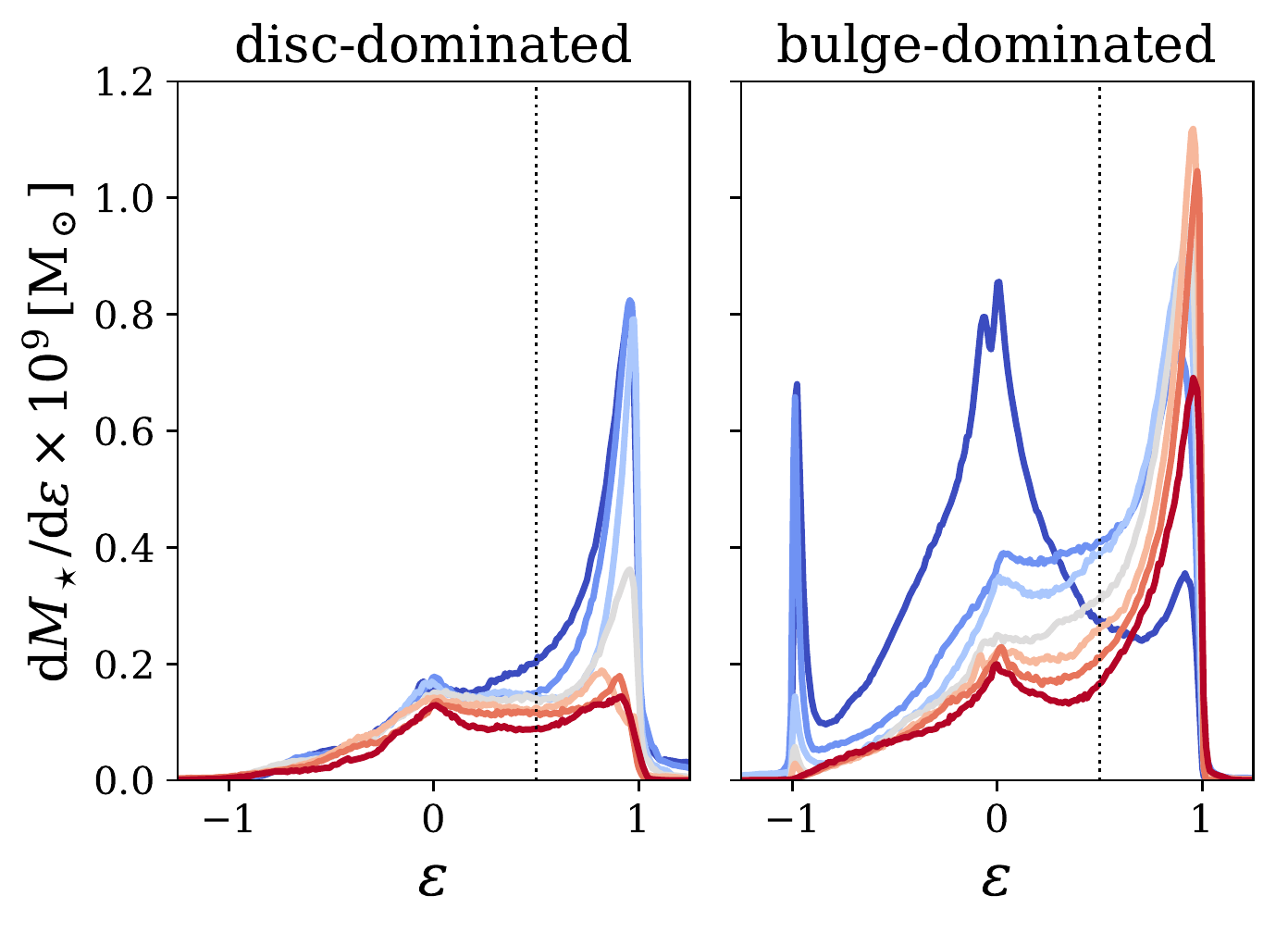}
    \caption{Evolution of the distribution of circularity $\epsilon$. Shown is the stellar mass per bin. The first snapshot in red coincides with the last star-burst: The disc-dominated galaxy (left) has similar mass in the bulge and disc-components. The bulge-dominated has a more dominant disc component. 
    Time evolution is shown as transitions to bluer colors in steps of $\simeq 1.5$ Gyrs until the final snapshot at $z=0$ in dark blue.
    Here the disc-dominated has a dominant disc-component and a small bulge. The bulge-dominated has a large bulge-component and counter-rotating stars.}
    \label{fig:circularity}
\end{figure}

\autoref{fig:circularity} shows the evolution of the circularity distribution at different epochs, 
uniformly spread between the last starburst and the present day $z=0$, in steps of $\simeq 1.5$~Gyr. The disc-dominated galaxy right after the last starburst shows a clear bulge component centred on $\epsilon=0$ and a small disc component. 

Following the (ad-hoc) convention described in the literature \citep[e.g.][]{2003ApJ...597...21A,2018MNRAS.473.1930E,2018MNRAS.477.4915O,2019ApJ...883...25P},
we use a fixed cutoff at $\epsilon=0.5$ (shown as a vertical dotted line in \autoref{fig:circularity}) 
to separate bulge stars from disc stars. The time evolution in \autoref{fig:circularity} reveals that the stellar disc component in the disc dominated system steadily grows between the
last starburst and the final epoch, with a three-fold increase in disc mass at an almost constant bulge mass. This level of disc growth, at constant bulge mass, is in excellent agreement with expected $L_\star$ galaxy evolution since $z\sim 1-1.5$ \citep[][]{2013ApJ...771L..35V}. 
Furthermore, this confirms recent studies which highlighted that at high redshift, galaxies were bulge-dominated and only in the last $\sim 10$Gyr started to form strong disc components \citep{2019ApJ...883...25P,2020MNRAS.491.3461B}.
During this last epoch from $z=2$ to $z=0$, the stellar mass of the system grew from $1.2 \times 10^{10} M_\odot$ to $3.4 \times 10^{10} M_\odot$ by a factor of $2.7$, the radius from $0.5$kpc to $3.1$kpc by a factor of $5.8$ and the rotation velocity from $94$km/s to $144$km/s by a factor $1.5$.
As a result the total stellar AM increased by a factor of $\sim 24$ \citep{2017MNRAS.467.3140S,2020MNRAS.495L..42R,2020MNRAS.491L..51P}.
For the bulge-dominated galaxy, on the other hand, the disc component centred at $\epsilon=+1$ does not grow significantly between
the last starburst and the present epoch, while the bulge component sees its mass grow by a factor of 2. Furthermore, we measure a spectacular rise of 
a counter-rotating stellar disc centred at $\epsilon=-1$, whose origin is directly related to the gaseous filament with negative contribution to the galaxy's angular momentum content
identified in \autoref{fig:am}; new stars form in the counter-rotating
disc that emerged from accretion of counter-rotating gas. 

Another spectacular feature in \autoref{fig:circularity} can be found for the last considered instance, where a dramatic evolution of the circularity distribution in the last 1.5~Gyr occurs due to a minor merger. Indeed, a satellite galaxy with stellar mass $M_\star \simeq 10^9 \mathrm{M}_\odot$ on a nearly radial orbit
hits the centre of our bulge-dominated galaxy, leading to a strong orbital reconfiguration of the disc stars into more eccentric orbits,
ultimately increasing significantly the bulge mass \citep{2013IAUS..295..340N, 2017ARA&A..55...59N}.

In summary, the disc-dominated galaxy emerges as a large disc because of the consistent (over several Gyr) and smooth accretion of large angular momentum material. Using the above mentioned disc-bulge decomposition, the post starburst disc-to-total ratio is $D/T=0.37$ and evolves to $D/T=0.61$ at $z=0$. In contrast the bulge-dominated galaxy suffers from multiple and combined destructive events, leading to the formation of a counter-rotating disc and a massive bulge. For this galaxy the post starburst $D/T=0.58$, but evolves to $D/T=0.19$ at $z=0$. We reiterate that the $z=0$ dark matter halo spin parameters are almost identical in the two galaxies, illustrating the fact that $\lambda$ is a poor predictor of the final galaxy morphology \citep[e.g.][]{2019MNRAS.488.4801J}.

Note that our bulge-dominated galaxy, which features a gas-rich, star forming nuclear disc, would be an ideal environment to grow a supermassive black hole and trigger strong AGN feedback. This would change the conditions in the galaxy and probably results in a completely dead, gas-poor spheroidal galaxy.

\subsection{Evolution of the Star Formation Intensity}

\begin{figure*}
    \centering
    \includegraphics[width=0.4\linewidth]{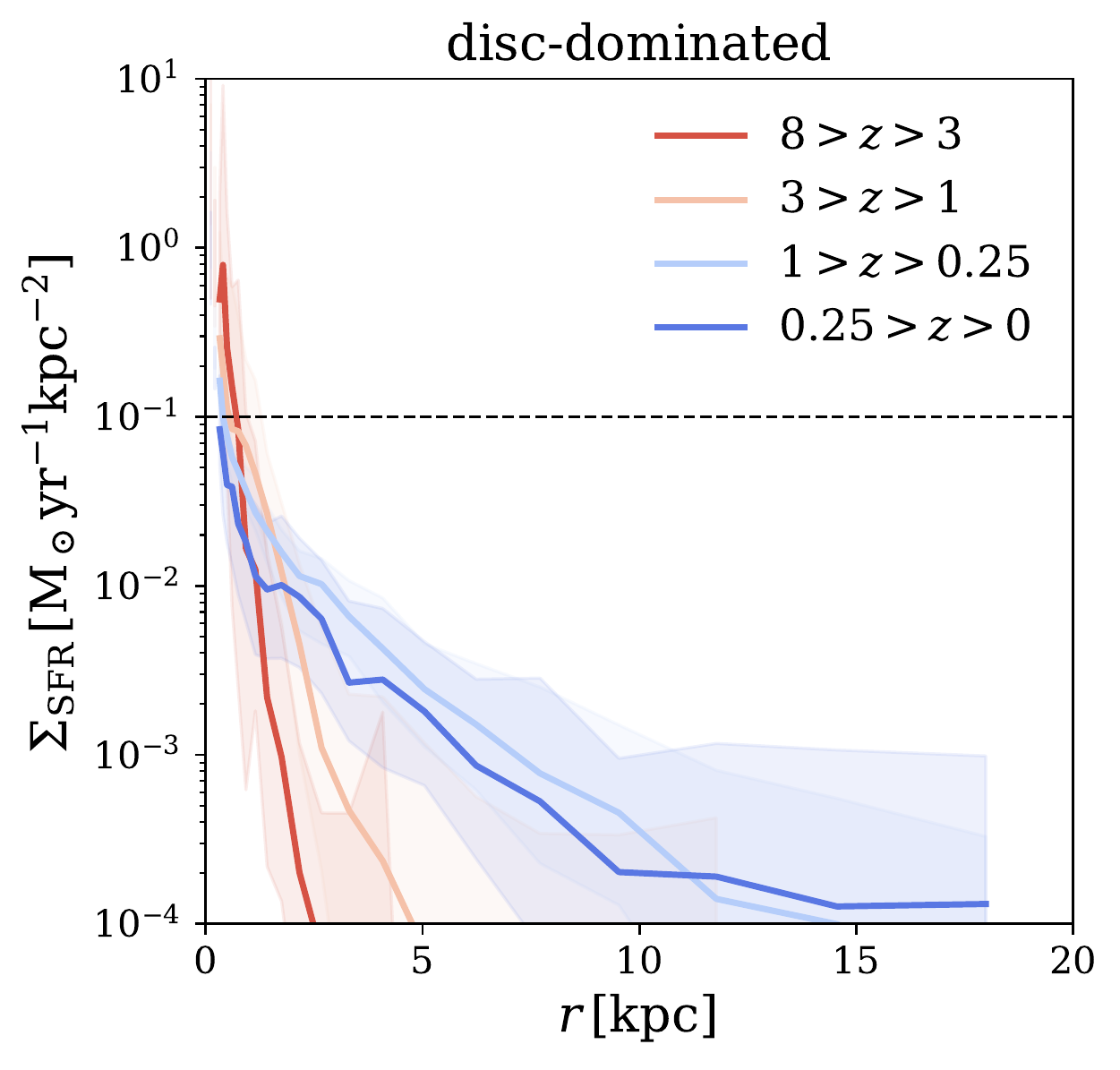}
    \includegraphics[width=0.4\linewidth]{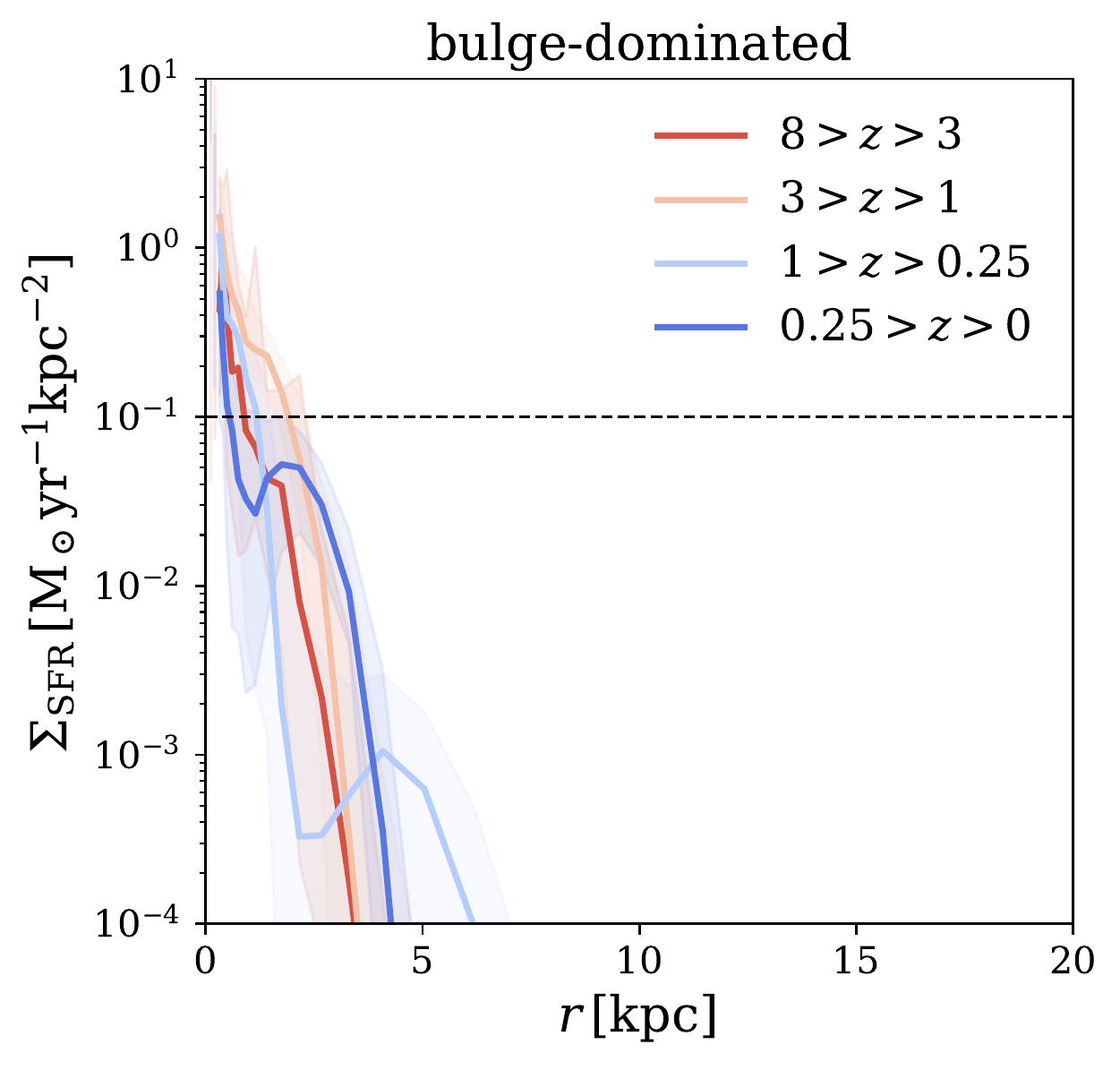}
    \caption{Radial profiles of the surface star-formation rate as a function of physical radius binned in different redshifts.
    }
    \label{fig:sig_sfr_r}
\end{figure*}

It is still unclear why the disc-dominated galaxy, after an early phase of strong starbursts and outflows, 
managed to form such a quiescent, thin disc of gas and stars that survived until today.
Other studies, based on different star formation and feedback implementations, have demonstrated that strong feedback, necessary at high redshift, leads to thick and disturbed discs at low redshift \citep{2011MNRAS.410.1391A,2013MNRAS.436..625S,2014MNRAS.444.2837R}. 
Moreover, inspecting the gas density maps of \autoref{fig:vis} shows
that the bulge-dominated galaxy has a significantly denser circum-galactic medium than the disc-dominated one, by almost a factor of ten.  

To interpret these results, we propose the following scenario, based on the time evolution of the 
surface density of star formation rate, also referred to as the star formation intensity (SFI), defined as
\begin{equation}
    \Sigma_\mathrm{SFR} = \frac{\mathrm{SFR}}{2 \pi r_\mathrm{SFR/2}^2},
    \label{eq:sig_sfr}
\end{equation}
where $r_\mathrm{SFR/2}$ is the half-SFR-radius.
In the last row of \autoref{fig:sfr_acc}, we show the evolution of $\Sigma_\mathrm{SFR}$ for both galaxies as a function of redshift.
We find that at high redshift both galaxies have high SFI with $\Sigma_\mathrm{SFR} \simeq 10 \,\mathrm{M_\odot yr^{-1} kpc^{-2}}$ 
maintained until $z \simeq 1.5$. After the Grand Twirl, however, the SFI in the disc-dominated galaxy drops very rapidly to a value smaller than 
$0.1\,\mathrm{M_\odot yr^{-1} kpc^{-2}}$ whereas the bulge-dominated galaxy maintains a high SFI until $z=0$.

The significant drop of the SFI in the disc-dominated case is naturally explained by 1) starburst quenching of star formation (reduced $\mathrm{SFR}$), due to rapid gas depletion and outflows, and 2) the rapid size evolution and emergence of an extended gas disc (increased $r_\mathrm{SFR/2}$).

Since our disc-dominated galaxy's key properties ($r_{e\star}, \mathrm{SFR}, M_\star$ and $\Sigma_\mathrm{SFR}$) as well as the resulting disc-to-total ratio at $z=0$ are similar to those obtained for the MW \citep[e.g.][found $D/T\sim0.6$]{2020MNRAS.494.4291C}, we are interested in comparing the evolution of the SFI of our simulations with observational data for the MW.
Although our disc-dominated galaxy might differ in certain quantities from the MW, such a comparison can shed light on the overall evolution of disc galaxies with respect to a possible Grand Twirl event.
We compare our results to the SFI of the Milky Way using the analysis performed by \cite{2014ApJ...789L..30L}
based on stars in the solar neighbourhood \citep{2013A&A...560A.109H,2014ApJ...781L..31S}, shown as a solid black
line in \autoref{fig:sfr_acc}. The agreement is striking, suggesting that a similar Grand Twirl phenomenon may have occurred in the Milky Way, around $z \simeq 1.5$.
Also shown in the figure is a range of characteristic values of $\Sigma_\mathrm{SFR}$ that have been proposed as critical thresholds for driving large-scale outflows. This value, may range from  $0.1\,\mathrm{M_\odot yr^{-1} kpc^{-2}}$ based on observations of nearby galaxies \citep{1996ApJ...472..546L, 2003RMxAC..17...47H} to $\sim1\,\mathrm{M_\odot yr^{-1} kpc^{-2}}$ at high redshift \citep{2012ApJ...761...43N,2019ApJ...875...21F}.
Our simulations support the same idea: the disc-dominated galaxy shows very little 
outflows after the last starburst, while the bulge-dominated galaxy continues to eject significant quantities of gas in the halo.
This explains the very different gas densities in the two galactic corona see in \autoref{fig:vis}. We will discuss this point
further in the next section.

The radial profiles of $\Sigma_\mathrm{SFR}$, shown in \autoref{fig:sig_sfr_r}, summarise and confirm this picture.
First, we find the rapid growth of the gas disc after the Grand Twirl of the disc-dominated case, while the bulge-dominated galaxy maintains a compact star-formation region.
We note that the central gas-rich, star forming nuclear disc of the bulge-dominated galaxy should probably host a supermassive black hole and trigger strong AGN feedback which are not included in our sub-grid physics models. Strong AGN feedback would probably turn the galaxy into a completely quenched and gas-poor spheroidal galaxy. However the main results of this paper concerning the importance of co-rotating accreted gas to form extended disc galaxies should not be affected by the lack of such a model.
Second, we see that the disc-dominated galaxy SFI drops below the critical threshold for starburst-driven outflows everywhere at late times, even in the central region, while the bulge-dominated galaxy retains regions above the threshold, leading to a consistent launching of outflows, even at late times. 

\section{Discussion}
\label{section:discussion}

We now discuss in more details the rapid growth of the large disc we have observed in our simulations, 
an episode that we called the Grand Twirl. We speculate this also happened in the Milky Way, as suggested by the analysis of \cite{2014ApJ...789L..30L}. We also discuss how this rapid disc growth sets the conditions for the emergence of a quiescent 
disc, in which supernovae-driven feedback becomes inefficient at launching outflows and at disturbing the thin disc. We note in passing
that our analysis in based only on two simulations and cannot be used to draw any statistical conclusions on the frequency of large discs 
predicted by current galaxy formation simulations \citep[for recent statistical studies, see e.g.][]{2014MNRAS.444.1453D,2015ApJ...804L..40G,2016MNRAS.460.4466Z,2017MNRAS.466.1625Z,2017ApJ...841...16D,2017MNRAS.464.3850L,2019ApJ...883...25P}

\subsection{The Emergence of Large discs}

The traditional picture of disc assembly is based on tidal torque theory, for which halos obtain their spin from their cosmic environment through tidal torques \citep{1969ApJ...155..393P, 1970Afz.....6..581D, 1984ApJ...286...38W}. Assuming that gas follows the dark matter dynamics on large-scales, emerging star forming discs are expected to have similar spins \citep{1980MNRAS.193..189F}.
However simulations with gas cooling and star formation have demonstrated that this simple picture is invalid. Several studies have shown that  the spin of baryons and dark matter inside the halo are barely correlated and that the two angular momentum distributions widely differ \citep{2002ApJ...576...21V,2015ApJ...812...29T, 2015ApJ...804L..40G,2019MNRAS.488.4801J,2017ApJ...843...47S}.

In both our simulations, we observe at high redshift ($z \simeq 1.5-2$) a spectacularly fast disc growth triggered by a combination of in-flowing pristine gas through cold streams and recycled metal enriched outflows that combined and quickly fueled the gas disc in the rapid event we called the Grand Twirl. This is in line with previous studies that highlighted the role played by cold stream accretion from the cosmic web of large-scale filaments in setting the spin of gaseous discs \citep{2011MNRAS.418.2493P,2014MNRAS.444.1453D}. Moreover, \cite{2013ApJ...769...74S} showed that the high angular momentum of gaseous disc observed in simulations is a direct consequence of this filamentary accretion. This cold stream accretion is the main component of the new theory of disc evolution proposed by \cite{2015MNRAS.449.2087D}.

The second important ingredient for the fast build-up of a large disc is the re-accretion of outflow material.
Indeed, feedback at these high redshifts is strong enough to remove low angular momentum gas from the disc and make it available in the halo to be efficiently torqued to large angular momentum by incoming cold streams.
This combination of pristine and recycled material with large angular momentum is then accreted quickly into this Grand Twirl. The same process was described in \cite{2014MNRAS.443.2092U} who argued that strong stellar feedback and re-accreted material can actually promote the formation of discs.

Comparing the later evolution of the discs in our two different halos, it appears crucial for the disc survival to have a sustained accretion of positive angular momentum gas that is added constructively to the disc. For the disc-dominated galaxy, the incoming cold gas from the cosmic web is almost exclusively co-rotating leading to the steady build-up of an extended gas disc. On the other hand, for the bulge-dominated galaxy, the incoming material is almost purely counter-rotating leading to the steady build-up of a large bulge.
The gas-rich and star forming nuclear disc would be an ideal environment to grow a supermassive black hole and trigger strong AGN feedback. The galaxy would probably end up as a passive bulge-dominated galaxy similar to those common at low redshift.

This picture is in agreement with previous studies that showed that the final galaxy morphology depends on the coherent alignment (or lack thereof) of the accreted angular momentum. In particular, \cite{2012MNRAS.423.1544S} compared the alignment of the angular momentum of baryons within a spherical shell measured at the turn-around radius with the total halo spin. The more consistently aligned the angular momentum in these shells is, the more likely a large disc will emerge at $z=0$. Our simulations confirm this interpretation but instead of considering the turn-around radius and the entire halo, we used instead the alignment of angular momentum within $R_{\rm vir}$ with the one of the central galaxy within $0.1 R_{\rm vir}$. This allows us to remove from the analysis satellites and gas streams within the halos that never reach the central region.

\subsection{The Emergence of Quiescent discs}

A spectacular consequence of the Grand Twirl is the emergence of a very quiescent disc, in striking contrast with the earlier epoch 
that was dominated by strong starbursts and associated outflows, resulting in very disturbed environments. 
After the rapid formation of the extended gas disc, the star formation surface density drops abruptly, in agreement with the model of \cite{2014ApJ...789L..30L} for the Milky Way. 

This is at odd with some earlier works for which strong feedback completely destroyed the final thin discs \citep{2012MNRAS.423.1726S,2013MNRAS.428..129S,2014MNRAS.444.2837R}, outlining an apparent contradiction between the strong feedback required 
at high redshift to regulate star formation and the weak feedback required at low redshift to allow the formation of quiescent and thin gas discs.
One issue in these earlier studies is the unrealistic modelling of stellar feedback. Indeed, instead of resolving successive individual supernovae (like we do here), these studies used individual star particle to represent entire star clusters and made them explode in one single energetic event.

In order to launch strong outflows, the star formation intensity has to be high enough, so that overlapping SN explosions create a volume-filling hot phase such that gas is removed in a thermally driven outflow. In the Milky Way and at the present epoch, the star formation intensity is too low and disfavour the build-up of overlapping hot bubbles and the formation of strong outflows \citep{2015MNRAS.449.1057G,2015ApJ...814....4L,2017ARA&A..55...59N,2020MNRAS.495.1035S}.
This has been demonstrated for example by \cite{2015ApJ...802...99K,2015MNRAS.454..238W, 2016MNRAS.456.3432G} using high-resolution simulations of stratified galactic discs with multiple SN explosions. These studies have also shown that the degree of clustering of SN explosions matters,  
with overlapping and percolating supernovae driving much stronger outflows  than individual, spatially separated events \citep{2017ApJ...834...25K, 2019MNRAS.483.3647G,2018MNRAS.481.3325F,2020MNRAS.492...79M}.
The Grand Twirl is the main mechanism causing this transition from highly clustered SN explosions driving strong outflows
to a more uniform distribution of isolated events unable to perturb the disc anymore. 

We believe these effects have been seen already in several past cosmological simulations of galaxy formation. For example, \cite{2015MNRAS.454.2691M} have computed the mass loading factor in several zoom-in simulations of disc galaxies, showing a striking and abrupt decline of the outflow rate
at late time and for Milky Way sized halos. Although this effect was attributed to the increased escape velocity in larger mass halos, 
we speculate that the rapid formation of a large disc must have also played an important role. \cite{2015ApJ...804...18A} modelled the formation of a Milky Way analogue and obtained also a rapid transition of the star formation rate, from very bursty at high redshift to very smooth at low redshift. The transition into an angular momentum rich disc occurred after the last major merger at $z>1$, and was likely due to a similar Grand Twirl episode.

\section{Conclusions}
\label{section:conclusions}
In this paper, we have investigated the evolution of two galaxies using hydrodynamical simulations of two halos that have masses similar to the estimated Milky Way halo mass at $z=0$. We have seen the emergence of two very different galaxies albeit their very similar halo properties. Our main results and conclusions can be summarised as follows:
\begin{itemize}
    \item Our simulated galaxies undergo an epoch of rapid disc growth at $z\simeq 1.5$, where the gas half-mass radius doubles in less than a Gyr. We called this event the Grand Twirl.
    During this epoch, both pristine gas from cold streams and re-accreted outflow material combine to form an extended gas disc in a very short time.
    \item For one galaxy, an extended and thin disc forms and persists until $z=0$, as the result of the consistent accretion of cold gas with positive angular momentum, adding constructively material to the initial disc that formed after the Grand Twirl episode. For the other galaxy, accreted gas with negative angular momentum favours the growth of the bulge component.
    \item The build up of this extended gas disc explains the transition from a high-redshift, outflow-dominated era to a very quiescent epoch at low redshift, for which feedback is not capable of launching outflows anymore.
\end{itemize}

For our disc-dominated galaxy, the time evolution of the surface star-formation density shows remarkable agreement with data obtained for the Milky Way. We speculate that the Milky Way may have undergone such a Grand Twirl episode at $z\simeq 1.5$.
This echoes with recent work speculating a similar massive accretion event at the origin of a peculiar structure in the metal enrichment properties of stars in the Milky Way as seen by Gaia \citep{2020MNRAS.494.3880B}. 
Star formation in such a rapidly formed outer disc could favour the emergence of a characteristic chemically bimodality in stellar abundance ratios between $\alpha$-elements and iron ([$\alpha$/Fe]) over a wide range of [Fe/H]. This is the focus of a follow-up project 
\citep{2020arXiv200606008A,2020arXiv200606011R,2020arXiv200606012R}.
In future work, we would also like to perform the same analysis presented here over a larger sample of halos.
This will allow us to determine whether such Grand Twirl episodes are a common feature in the past history of large galactic halos.

\section*{Acknowledgements}
The authors thank the referee for their constructive
comments that improved the quality of the paper.
We acknowledge stimulating discussions with Lorenzo Posti, Pedro Capelo, Lucio Mayer and Robert Feldmann.
This work was supported by the Swiss National Supercomputing Center (CSCS) project s890 - ``Predictive models for galaxy formation'' 
and the Swiss National Science Foundation (SNSF) project 172535 - ``Multi-scale multi-physics models of galaxy formation".
The simulations in this work were performed on Piz Daint at the Swiss Supercomputing Center (CSCS) in Lugano,
and the analysis was performed with equipment maintained by the Service and Support for Science IT, University of Zurich.
We made use of the pynbody package \citep{2013ascl.soft05002P}. OA acknowledges support from the Knut and Alice Wallenberg Foundation and the Swedish Research Council (grant 2014-5791).

\section*{Data Availability}
The data underlying this article will be shared on reasonable request to the corresponding author.
%
%%%%%%%%%%%%%%%%%%%%%%%%%%%%%%%%%%%%%%%%%%%%%%%%%%
%
%%%%%%%%%%%%%%%%%%%% REFERENCES %%%%%%%%%%%%%%%%%%
%
% The best way to enter references is to use BibTeX:
%
\bibliographystyle{mnras}
\bibliography{export}
%
%%%%%%%%%%%%%%%%%%%%%%%%%%%%%%%%%%%%%%%%%%%%%%%%%%
%
%%%%%%%%%%%%%%%%% APPENDICES %%%%%%%%%%%%%%%%%%%%%
%
%\appendix
%%%%%%%%%%%%%%%%%%%%%%%%%%%%%%%%%%%%%%%%%%%%%%%%%%
% Don't change these lines
\bsp	% typesetting comment
\label{lastpage}
\end{document}